\documentclass[aps,prl,twocolumn,nopacs,superscriptaddress]{revtex4}
\usepackage[utf8]{inputenc}

\usepackage{graphicx}  
\usepackage{dcolumn}   
\usepackage{bm}        
\usepackage{amssymb}   
\usepackage{amsmath}
\usepackage{units}
\usepackage[english]{babel}
\usepackage[dvipsnames]{xcolor}
\usepackage{xfrac}

\usepackage[%
colorlinks=true,
urlcolor=RoyalBlue,
linkcolor=RoyalBlue,
citecolor=RoyalBlue,
]{hyperref}

\usepackage{sidecap}
\sidecaptionvpos{figure}{t}  

\usepackage[type1]{libertine}                                        
\usepackage{textcomp}
\usepackage[scaled=.85]{beramono}
\usepackage[libertine,cmintegrals,cmbraces,vvarbb,slantedGreek]{newtxmath}
\usepackage[scr=boondoxo]{mathalfa}
\usepackage{bm}
\usepackage[lf]{carlito}

\usepackage{braket}
\usepackage{bm}

\makeatletter
\DeclareRobustCommand{\cev}[1]{%
  \mathpalette\do@cev{#1}%
}
\newcommand{\do@cev}[2]{%
  \fix@cev{#1}{+}%
  \reflectbox{$\m@th#1\vec{\reflectbox{$\fix@cev{#1}{-}\m@th#1#2\fix@cev{#1}{+}$}}$}%
  \fix@cev{#1}{-}%
}
\newcommand{\fix@cev}[2]{%
  \ifx#1\displaystyle
    \mkern#23mu
  \else
    \ifx#1\textstyle
      \mkern#23mu
    \else
      \ifx#1\scriptstyle
        \mkern#22mu
      \else
        \mkern#22mu
      \fi
    \fi
  \fi
}
\makeatother

\newcommand{\spinhalf}{spin-\sfrac{1}{2}}
\renewcommand{\Re}{\text{Re}}
\renewcommand{\Im}{\text{Im}}

\begin{document}

\title{Universal scaling of tunable Yu-Shiba-Rusinov states across the quantum phase transition}

\author{Haonan Huang}
\affiliation{Max-Planck-Institut f\"ur Festk\"orperforschung, Heisenbergstraße 1,
70569 Stuttgart, Germany}
\author{Sujoy Karan}
\affiliation{Max-Planck-Institut f\"ur Festk\"orperforschung, Heisenbergstraße 1,
70569 Stuttgart, Germany}
\author{Ciprian Padurariu}
\affiliation{Institut für Komplexe Quantensysteme and IQST, Universität Ulm, Albert-Einstein-Allee 11, 89069 Ulm, Germany}
\author{Bj\"orn Kubala}
\affiliation{Institut für Komplexe Quantensysteme and IQST, Universität Ulm, Albert-Einstein-Allee 11, 89069 Ulm, Germany}
\affiliation{Institute of Quantum Technologies, German Aerospace Center (DLR), S\"oflinger Stra{\ss}e 100, 89077, Ulm, Germany}
\author{Juan Carlos Cuevas}
\affiliation{Departamento de F\'{\i}sica Te\'orica de la Materia Condensada and
Condensed Matter Physics Center (IFIMAC), Universidad Aut\'onoma de Madrid, 28049 Madrid, Spain}
\author{Joachim Ankerhold}
\affiliation{Institut für Komplexe Quantensysteme and IQST, Universität Ulm, Albert-Einstein-Allee 11, 89069 Ulm, Germany}
\author{Klaus Kern}
\affiliation{Max-Planck-Institut f\"ur Festk\"orperforschung, Heisenbergstraße 1,
70569 Stuttgart, Germany}
\affiliation{Institut de Physique, Ecole Polytechnique Fédérale de Lausanne, 1015 Lausanne, Switzerland}
\author{Christian R. Ast}
\email[Corresponding author; electronic address:\ ]{c.ast@fkf.mpg.de}
\affiliation{Max-Planck-Institut f\"ur Festk\"orperforschung, Heisenbergstraße 1,
70569 Stuttgart, Germany}

\date{\today}

\begin{abstract}
Quantum magnetic impurities give rise to a wealth of phenomena attracting tremendous research interest in recent years. On a normal metal, magnetic impurities generate the correlation-driven Kondo effect. On a superconductor, bound states emerge inside the superconducting gap called the Yu-Shiba-Rusinov (YSR) states. Theoretically, quantum impurity problems have been successfully tackled by numerical renormalization group (NRG) theory, where the Kondo and YSR physics are shown to be unified and the normalized YSR energy scales universally with the Kondo temperature divided by the superconducting gap. However, experimentally the Kondo temperature is usually extracted from phenomenological approaches, which gives rise to significant uncertainties and cannot account for magnetic fields properly. Using scanning tunneling microscopy at 10\,mK, we apply a magnetic field to several YSR impurities on a vanadium tip to reveal the Kondo effect and employ the microscopic single impurity Anderson model with NRG to fit the Kondo spectra in magnetic fields accurately and extract the corresponding Kondo temperature unambiguously. Some YSR states move across the quantum phase transition (QPT) due to the changes in atomic forces during tip approach, yielding a continuous universal scaling with quantitative precision for quantum \spinhalf\ impurities.  
\end{abstract}

\maketitle

\section*{Introduction}

Quantum impurity problems deal with impurities featuring a small number of intrinsic quantum degrees of freedom connected to a continuous bath. The complexity lies in the quantum nature of such impurities resulting in correlations and many-body effects as well as the concomitant breakdown of a perturbative or mean field treatment \cite{bulla2008numerical}. Magnetic impurities are particularly interesting as it is not \textit{a priori} clear whether a fully quantum theory is necessary or a semiclassical treatment on the spin degrees of freedom suffices to describe the system. In some instances, the spin can be treated classically, as in the case of the in-gap Yu-Shiba-Rusinov (YSR) states generated from magnetic impurities on superconducting surfaces \cite{Yu1965,Shiba1968,Rusinov1969,Yazdani1997}. There, correlation effects play a subordinate role and, therefore, the YSR states can be treated well within the mean field approximation \cite{Balatsky2006,Menard2015,Ruby2015a,Ruby2018,Kezilebieke2018,Huang2020tunneling,Villas2020,villas2021tunneling,huang2020spin,karan2021superconducting}. However, in other cases the spin needs to be treated fully quantum mechanically and correlation effects are indispensable to describe the observations, such as the Kondo effect, which emerges when magnetic impurities are placed on normal conducting surfaces \cite{kondo1964resistance,hewson1997kondo,li1998kondo,madhavan1998tunneling,otte2008role,Ternes2009}.

The only difference between YSR and Kondo physics is a superconducting instead of a normal conducting bath. Thus, both phenomena can be theoretically described in a unified framework within the single impurity Anderson model (SIAM). A solution of such a model including full correlations can be achieved by means of numerical renormalization group (NRG) theory \cite{wilson1975renormalization,krishna1980renormalization_a,krishna1980renormalization_b}. The hallmark of this theory is a universal scaling relation between the normalized YSR energy $\varepsilon_\text{YSR}/\Delta$ and $k_\text{B}T_\text{K}/\Delta$, where $\varepsilon_\text{YSR}$ is the YSR energy, $T_\text{K}$ is the Kondo temperature, $k_\text{B}$ is the Boltzmann constant and $\Delta$ is the superconducting order parameter. The quantum phase transition (QPT) associated with YSR-physics is predicted to occur at $k_\text{B}T_\text{K}/\Delta=0.24$ \cite{Yoshioka2000,bauer2007spectral,bulla2008numerical,kadlecova2019practical,supinf}.

Despite these convincing theoretical predictions for the universal scaling, there are different aspects of uncertainty in existing experimental results, especially in atomic junctions where fitting the tunneling spectra is essential for extracting the Kondo temperature \cite{Franke2011,bauer2013microscopic,odobesko2020observation,lee2017scaling,kamlapure2019investigation}. One reason is that the commonly reported Kondo systems are not purely \spinhalf, which complicates the analysis. Another reason is that the experimental procedures followed for the extraction of the Kondo temperature are usually phenomenological, either done by simply taking the half width half maximum (HWHM) of the Kondo peak or by fitting with rather phenomenological functions like Fano or Frota functions \cite{Ternes2009,gruber2018kondo}, while a much better solution is to employ NRG directly \cite{Luitz2012,lee2017scaling}. Furthermore, the finite experimental temperature results in a thermal broadening obscuring the spectral peak. These obstacles as well as the existence of various definitions of the Kondo temperature result in an ambiguity of the obtained Kondo temperature, up to a factor of four \cite{bauer2013microscopic,supinf}.

\begin{SCfigure*}[][ht]
    \centering
    \includegraphics[width=0.66\textwidth]{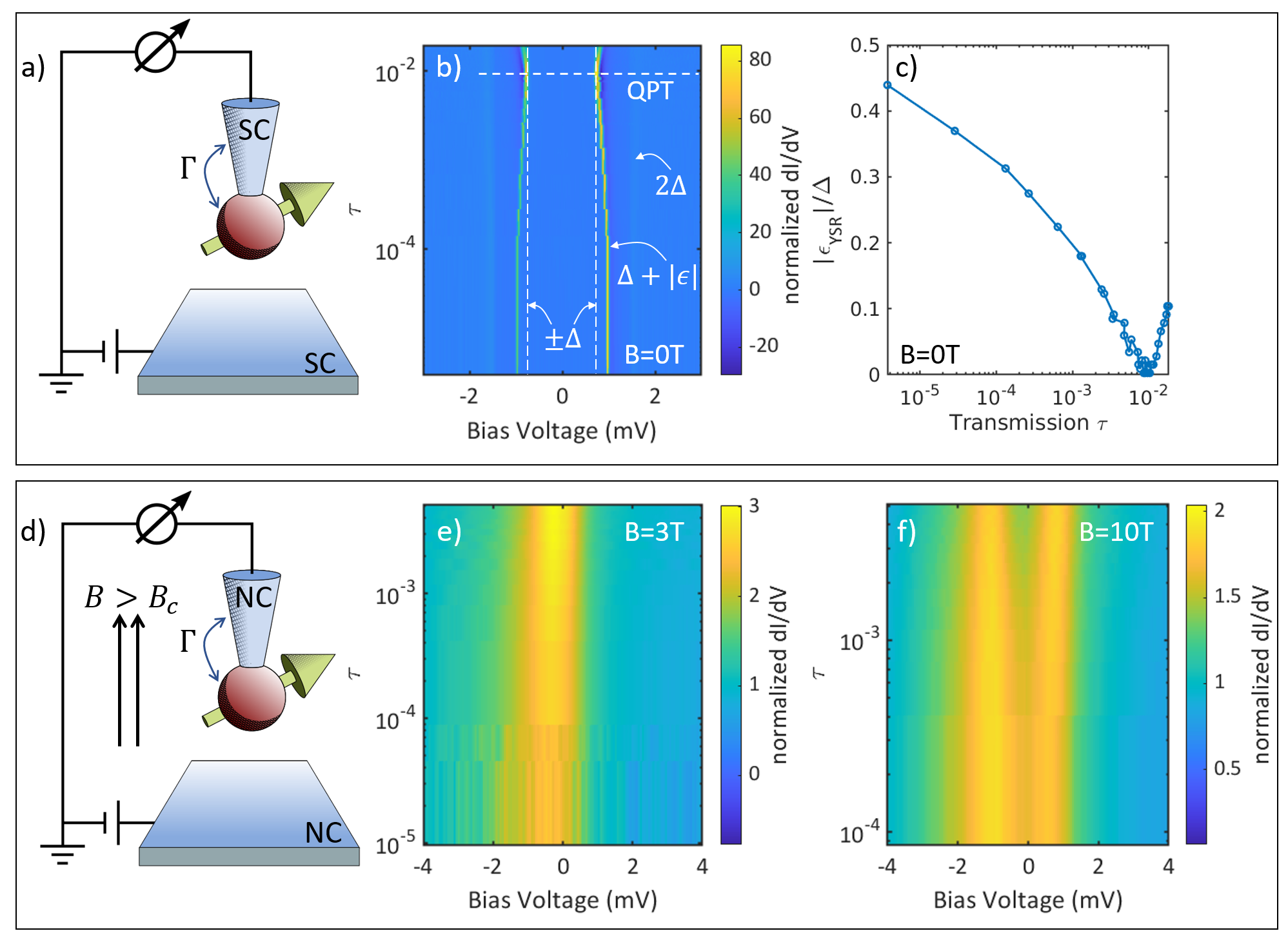}
    \caption{\textbf{A Yu-Shiba-Rusinov (YSR) impurity moving through the quantum phase transition (QPT).} (a) The schematics of the experiment at zero field, where the junction is superconducting. During tip approach, the atomic force exerted on the magnetic impurity changes resulting in a varying coupling $\Gamma$ and therefore a moving YSR energy. (b) YSR spectra across the QPT during tip approach. Here, since both tip and sample are superconducting, the spectral peaks involving the YSR states are at $\pm(\Delta+|\varepsilon_\text{YSR}|)$, where $\Delta$ is the superconducting order parameter and $\varepsilon_\text{YSR}$ is the energy of the tip YSR state. Therefore, the QPT happens when such spectral peaks touch $\pm\Delta$. (c) The extracted YSR energy normalized by $\Delta$ as a function of the transmission $\tau$. (d) The schematics of the experiment in a magnetic field higher than the critical magnetic field $B_c$, where superconductivity in both the tip and the sample are quenched. (e) Emerging Kondo peaks at 3\,T. (f) The Kondo peaks further split at high magnetic field ($B=10$\,T here). Note: the data presented in panels (b) and (c) are reproduced from Ref. \cite{karan2021superconducting}.}
    \label{fig1}
\end{SCfigure*}

Here, we follow a more rigorous procedure and extract the Kondo temperature unambiguously by employing the full NRG theory on the microscopic SIAM to directly fit the Kondo spectra on \spinhalf\ YSR impurities on the vanadium tip \cite{Huang2020tunneling,karan2021superconducting,huang2021phd} in magnetic fields at low temperatures ($10\,\text{mK}$). We compare the Kondo temperature with the YSR energy at zero field and demonstrates the universal scaling. Since both phenomena are treated within one microscopic framework, the ambiguity of the Kondo temperature is eliminated. Using this approach, we analyze various YSR impurities on the vanadium tip, especially those moving across the YSR-QPT with varying tip-sample distance, and reveal a continuous and quantitative universal scaling behavior.

\section*{Results}

\textbf{Quantum phase transition and the Kondo effect.} The magnetic impurity is introduced onto the apex of the vanadium scannning tunneling microscope (STM) tip producing a YSR state on the tip apex using the method described in Refs. \onlinecite{Huang2020tunneling} and \onlinecite{karan2021superconducting} (Fig.\ \ref{fig1}(a)). The sample is also vanadium, having the same order parameter $\Delta= 750\,\upmu$eV as the tip. At zero magnetic field and a base temperature of 10\,mK, the junction is well superconducting. The tunneling between the tip YSR state with the energy $\varepsilon_\text{YSR}$ and the coherence peak of the sample at $\Delta$ yields the most prominent spectral feature shown in Fig.\ \ref{fig1}(b) at bias voltages of $eV=\pm(\Delta+|\varepsilon_\text{YSR}|)$ (YSR-BCS peaks). The tunneling between the coherence peaks (BCS-BCS peaks) at bias voltages of $eV=\pm 2\Delta$ is largely suppressed compared to the YSR-BCS tunneling due to the presence of YSR states. Notice that from the YSR-BCS peak positions it is not possible to extract the sign of the YSR energy $\varepsilon_\text{YSR}$. We, therefore, use its absolute value in Figs.\ \ref{fig1} and \ref{fig2}. Conventionally, $\varepsilon_\text{YSR}$ acquires a negative sign after the QPT (in the screened spin regime) to indicate the fact that the ground state and the excited state interchange at the QPT. After we obtain more information with respect to the two domains of the QPT from Kondo spectra, we add the proper signs to $\varepsilon_\text{YSR}$ (Figs.\ \ref{fig3} and \ref{fig4}).

In this example, the YSR state moves across the QPT when approaching the tip to the sample and increasing the normal state transmission $\tau=G_N/G_0$, where $G_N$ is the normal state conductance measured well above the superconducting gap and $G_0=2e^2/h$ is the quantum of conductance ($e$ is electron charge and $h$ is Planck constant). The clear indication of this QPT is the zero crossing of $\varepsilon_\text{YSR}$ signaled by the point where the YSR-BCS peaks touch $eV = \pm\Delta$ (dashed vertical line in Fig.\ \ref{fig1}(b)). We further find $|\varepsilon_\text{YSR}|$ by subtracting $\Delta$ from the YSR-BCS peak position and plot the normalized YSR energy $|\varepsilon_\text{YSR}|/\Delta$ in Fig.\ \ref{fig1}(c), which clearly shows the zero crossing.

Then we apply magnetic fields from 1.5\,T to 10\,T perpendicular to the sample surface, which exceed the critical field and quench superconductivity in both the tip and the sample making them both normal conducting (Fig.\ \ref{fig1}(d)). At a magnetic field of 3\,T, a Kondo peak is observed around zero bias voltage (Fig.\ \ref{fig1}(e)), whose width decreases with increasing transmission, indicating a decreasing Kondo temperature. Increasing the magnetic field to 10\,T, the Kondo peak splits into two peaks due to the Zeeman effect (Fig.\ \ref{fig1}(f)). 

\textbf{The single impurity Anderson model and the universal scaling.} For a quantitative modeling, we use the SIAM with the Hamiltonian $H_\text{SIAM}=H_\text{s}+H_\text{i}+H_\text{c}$, where $H_\text{s}=\sum_{k\sigma}\xi_k c^\dagger_{k\sigma}c_{k\sigma}$ represents the band dispersion of the normal conducting substrate with $c$ and $c^\dagger$ being the annihilation and creation operators for electrons with momentum $k$ and spin $\sigma$, $H_\text{i}=\varepsilon_d(n_{\uparrow}+n_{\downarrow})+Un_{\uparrow}n_{\downarrow}$ denotes the Anderson impurity with impurity level $\varepsilon_d$, on-site Coulomb term $U$ and the electron occupation number operator $n$, and $H_\text{c}=\sum_{k\sigma}V_k(c^\dagger_{k\sigma}d_{\sigma}+\text{H.c.})$ represents the coupling between the impurity and the substrate with $d$ being the annihilation operator on the impurity site \cite{anderson1961localized,zitko2007many}. Assuming a constant density of states $\rho$ and constant hybridization strength $V_k$ near the Fermi level, we further define the parameter $\Gamma=\pi\rho|V_{k_\text{F}}|^2$ quantifying the impurity-substrate coupling. Therefore, the behavior of the SIAM is essentially defined by the three parameters $U$, $\varepsilon_d$ and $\Gamma$ (Fig.\ \ref{fig2}(a)). Note that all parameters are energies in units of the half bandwidth $D$ of the bulk conduction electrons. Among these three parameters, $\Gamma$ is of special experimental interest, because the movement of the YSR state shown in Fig.\ \ref{fig1}(b) originates from a changing impurity-substrate coupling $\Gamma$ due to the changing atomic forces acting in the junction between tip and sample \cite{Huang2019magnetic,karan2021superconducting}. Therefore, we will use $\Gamma$ as the variable to fit the Kondo spectra in the following. For convenience, we also define the asymmetry parameter $\delta=\varepsilon_d+U/2$ to be used instead of $\varepsilon_d$ as it is easier to distinguish different scenarios. If $\delta=0$, the impurity levels are symmetric around the Fermi energy resulting in electron-hole symmetry, yielding a symmetric spectral function. If $\delta\neq 0$, the spectral function will be asymmetric due to electron-hole asymmetry.

Through a Schrieffer-Wolff transformation \cite{schrieffer1966relation,kadlecova2019practical}, an expression of the Kondo temperature in terms of the SIAM parameters can be derived as 
\begin{equation}
\label{eq_kondoT}
    k_\text{B}T_\text{K} = D_\text{eff}\sqrt{\rho J}\exp\left(-\frac{1}{\rho J}\right)\text{, with } \rho J=\frac{8\Gamma}{\pi U}\frac{1}{1-4(\delta/U)^2}.
\end{equation}
The effective band width $D_\text{eff}$ depends on $U$: If $U\ll 1$, $D_\text{eff}=0.182U\sqrt{1-4(\delta/U)^2}$ \cite{Yoshioka2000,zitko2007many}. If $U\gg 1$, $D_\text{eff}$ is a constant on the order of 1, whose exact value is fixed comparing with the situation $U\ll 1$. In the NRG implementation used in this paper $D_\text{eff} = 0.5$ for $U\gg 1$ \cite{supinf}.

\begin{figure}[t]
    \centering
    \includegraphics[width=\columnwidth]{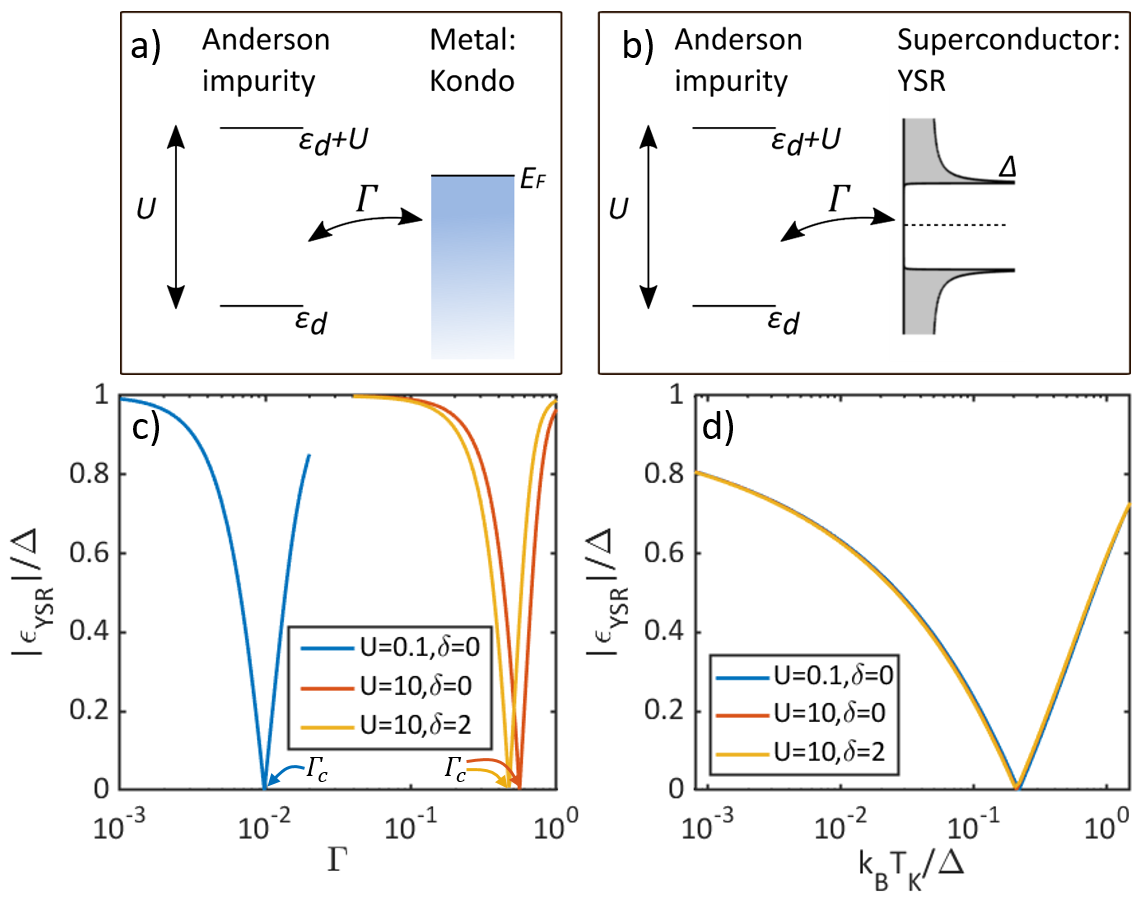}
    \caption{\textbf{Single impurity Anderson model (SIAM).} (a) The schematics of the SIAM for a normal metal giving rise to the Kondo effect. The parameters are the impurity level $\varepsilon_d$, the on-site Coulomb $U$ and the impurity-substrate coupling $\Gamma$. For convenience, in this paper we use the asymmetry parameter $\delta=\varepsilon_d+\frac{U}{2}$ instead of $\varepsilon_d$. (b) The schematics of the SIAM including superconductivity giving rise to the YSR states. Here, there is one more parameter in the model which is the superconducting order parameter $\Delta$. (c) The normalized YSR energy $|\varepsilon_\text{YSR}|/\Delta$ as a function of $\Gamma$ calculated from the numerical renormalization group (NRG) theory. The critical $\Gamma$ of the QPT ($\Gamma_\text{c}$) depends on both $U$ and $\delta$. (d) The normalized YSR energy as a function of $k_\text{B}T_\text{K}/\Delta$ is a universal curve independent of $U$ and $\delta$, indicating that the low energy physics is controlled by the Kondo temperature making not all parameters in the SIAM relevant.}
    \label{fig2}
\end{figure}

Further, to reflect the Kondo effect in a finite magnetic field (Figs.\ \ref{fig1}(d-f)), a Zeeman term needs to be added in the Hamiltonian
\begin{equation}
    H_\text{SIAM,B}=H_\text{SIAM}+g\mu_\text{B} BS_z,
\end{equation}
where $g$ is the gyromagnetic ratio, $\mu_\text{B}$ is the Bohr magneton, $S_z$ is the spin component along the magnetic field $B$.

On the other hand, to account for superconductivity and YSR states (Figs.\ \ref{fig1}(a-c)), the Hamiltonian is modified to 
\begin{equation}
    H_\text{SIAM,SC}=H_\text{SIAM}-\sum_{k}\left(\Delta c^\dagger_{k\uparrow}c^\dagger_{-k\downarrow}+\Delta^*c_{-k\downarrow}c_{k\uparrow}\right),
\end{equation}
with $\Delta$ being the superconducting order parameter (Fig.\ \ref{fig2}(b)). Using the ``NRG Ljubljana'' package for NRG simulations \cite{pruschke2009energy}, we calculated the normalized YSR energy $|\varepsilon_\text{YSR}|/\Delta$ as a function of $\Gamma$ for different $U$ and $\delta$ (Fig.\ \ref{fig2}(c)). With increasing $\Gamma$, the YSR energy always moves from the gap edge towards zero energy, across the QPT, and then from zero energy towards the gap edge again. The critical impurity-substrate coupling $\Gamma_\text{c}$ at the QPT depends on both $U$ and $\delta$.

We convert $\Gamma$ to $k_\text{B}T_\text{K}$ using Eq.\ \eqref{eq_kondoT} and plot $k_\text{B}T_\text{K}/\Delta$ as the horizontal axis in Fig.\ \ref{fig2}(d). All curves from Fig.\ \ref{fig2}(c) overlap now, demonstrating the universal scaling between the YSR energy and the Kondo temperature independent of the details of the SIAM parameters $\delta$ and $U$. This is because the low energy physics including the YSR and Kondo phenomena is dominated by the Kondo temperature, which is much smaller than the energy scale of $U$, $\varepsilon_d$ and $\Gamma$. Therefore, according to Eq.\ \eqref{eq_kondoT} there are some redundant degrees of freedom in the SIAM in the context of the Kondo effect and YSR states. The parameter $\delta$ is responsible for the asymmetry of the spectrum and remains thus relevant for fitting, but $U$ can be chosen freely. In the following, we use $U=10$, which is much larger than one, so that the impurity Hubbard satellite peaks, which are generally not observed in our experimental spectra, do not interfere with the fitting \cite{supinf}.

\textbf{Fitting the Kondo spectra directly using NRG theory.} The Kondo spectra from Figs.\ \ref{fig1}(e) and (f) are replotted as cascaded curves in Figs.\ \ref{fig3}(a) for $B=3$\,T and (b) for $B=10$\,T. We can directly fit the Kondo spectra in the magnetic field using NRG theory (black dashed lines in Figs.\ \ref{fig3}(a) and (b)). For the fit, we fix $U=10$, and also fix a $\delta$ that represents the asymmetry of the spectra for a certain impurity. For the impurity shown in Figs.\ \ref{fig1} and \ref{fig3}, we choose $\delta=-2$.

\begin{figure}
    \centering
    \includegraphics[width=\columnwidth]{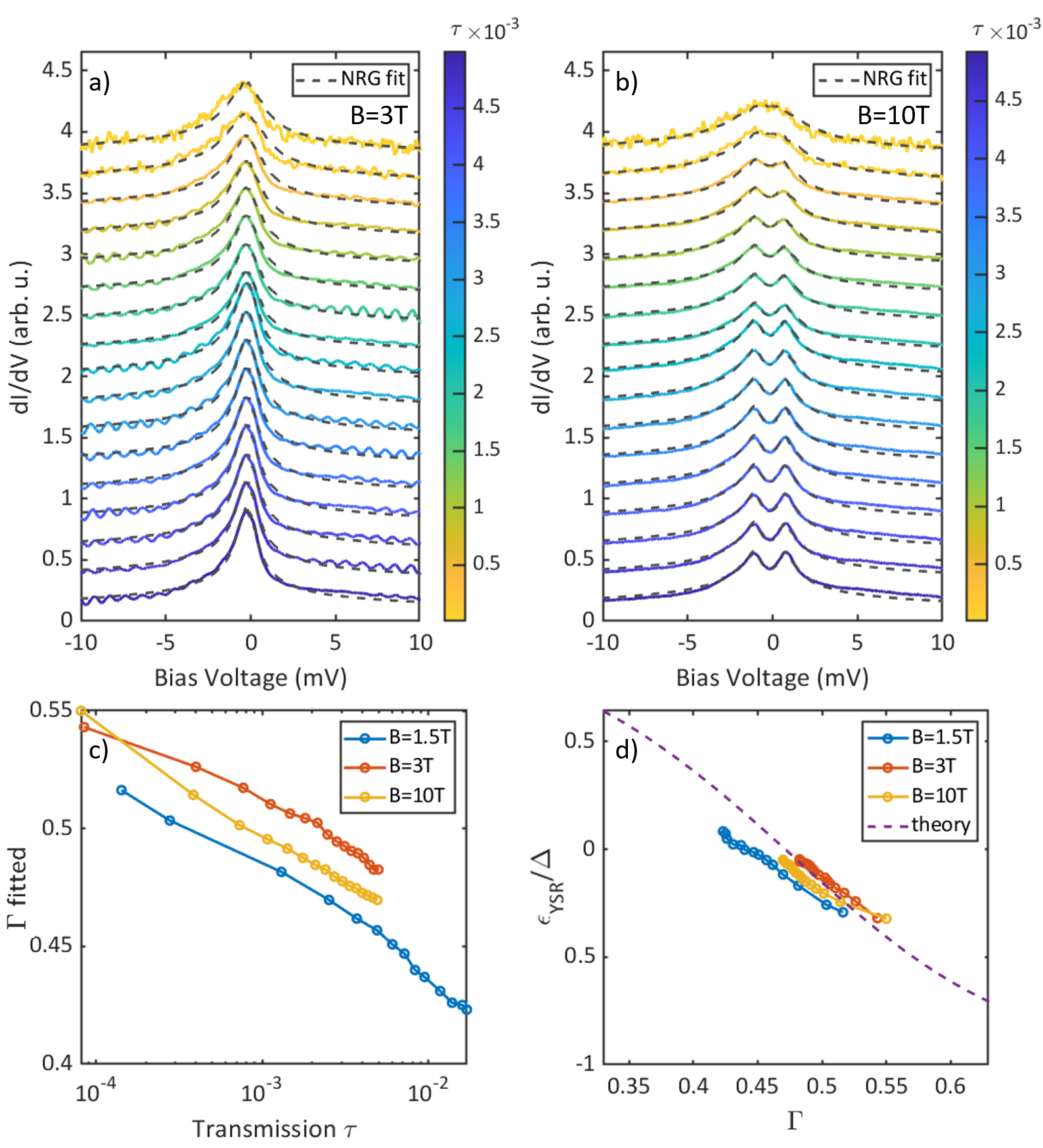}
    \caption{\textbf{Fitting the Kondo spectra of a YSR impurity moving through the QPT using the microscopic NRG theory.} (a) The Kondo spectra (colored solid lines) and their fits (black dashed lines) at $B=3$\,T. (b) The Kondo spectra (colored solid lines) and their fits (black dashed lines) at $B=10$\,T. (c) The fitted $\Gamma$ as a function of $\tau$ shows a decreasing coupling during tip approach, indicating that the impurity is pulled away due to attractive forces in the junction. (d) Normalized YSR energy as a function of $\Gamma$ extracted from Kondo peaks at $B=1.5$\,T (blue curve), 3\,T (red curve) and 10\,T (yellow curve). The purple dashed line is the theoretical expectation from the NRG theory with the same $\delta=-2$ as in the fits. Notice that the deviation in terms of $\Gamma$ is quite small (below 5\%).}
    \label{fig3}
\end{figure}

The asymmetry of a Kondo peak is typically ascribed to a Fano line shape originating from the interference between different tunneling paths \cite{ujsaghy_theory_2000,frota_shape_1992,fano_effects_1961,pruser_long-range_2011,farinacci_interfering_2020}. However, independent from these scattering and transport phenomena, an electron-hole asymmetry will also result in an asymmetric Kondo peak. The same electron-hole asymmetry will render the YSR peaks asymmetric, so a correlation between the asymmetry of a Kondo peak and the asymmetry of the corresponding YSR states is expected. In our experiment, the tunneling channel to the continuum is largely suppressed as evidenced by the faint BCS-BCS tunneling peaks in Fig.\ \ref{fig1}(b). Consequently, we conclude that the transport is dominantly through a single YSR channel, and thus Fano-like interference is minimal, which is also supported by the clearly peaked Kondo features in Figs.\ \ref{fig3}(a) and (b). A more detailed discussion of the alternative scenario considering Fano process is in the supplementary information \cite{supinf}, where we show that actually very similar Kondo temperatures are obtained for both scenarios \cite{zitko2011kondo}. For the following part here, however, we assume that the asymmetry of the Kondo peaks originates from the intrinsic asymmetry of the SIAM with parameter $\delta$. With $U=10$ and $\delta=-2$ being fixed as outlined above, the only fitting parameter is $\Gamma$. The resulting fits agree remarkably well with the experimental spectra, for both 3\,T (Fig.\ \ref{fig3}(a)), and 10\,T (Fig.\ \ref{fig3}(b)) magnetic fields. The general trend that the peak width reduces when increasing the normal state transmission $\tau$, the peak asymmetry, as well as the details of the Zeeman splitting at 10\,T are precisely reproduced.

We plot the fitted values for the impurity-substrate coupling $\Gamma$ as a function of transmission $\tau$ in Fig.\ \ref{fig3}(c). All fits for $B=1.5 $\,T, $3$\,T and $10$\,T are quite similar indicating consistency. The values also reflect the observation in the raw data (Figs.\ \ref{fig1}(e) and \ref{fig3}(a)) that the impurity-substrate coupling $\Gamma$ and concomitantly the Kondo temperature $T_\text{K}$ as well as the peak width decrease with increasing transmission.

Combining Fig.\ \ref{fig3}(c) and Fig.\ \ref{fig1}(c), we obtain the YSR energy $\varepsilon_\text{YSR}/\Delta$ as a function of the impurity-substrate coupling $\Gamma$, which is displayed in Fig.\ \ref{fig3}(d). The expected dependence of $\varepsilon_\text{YSR}/\Delta$ vs.\ $\Gamma$ predicted from the NRG model is shown as the purple dashed line. The experimental data generally follows the trend of the theoretical expectations and the critical impurity-substrate coupling $\Gamma_\text{c}$ at the QPT agrees approximately. The relative deviation in $\Gamma$ is quite small ($<5\%$), which could be due to slightly different atomic forces acting on the impurity when the substrate is superconducting or normal conducting, which will be discussed in more detail below.

\begin{figure}
    \centering
    \includegraphics[width=\columnwidth]{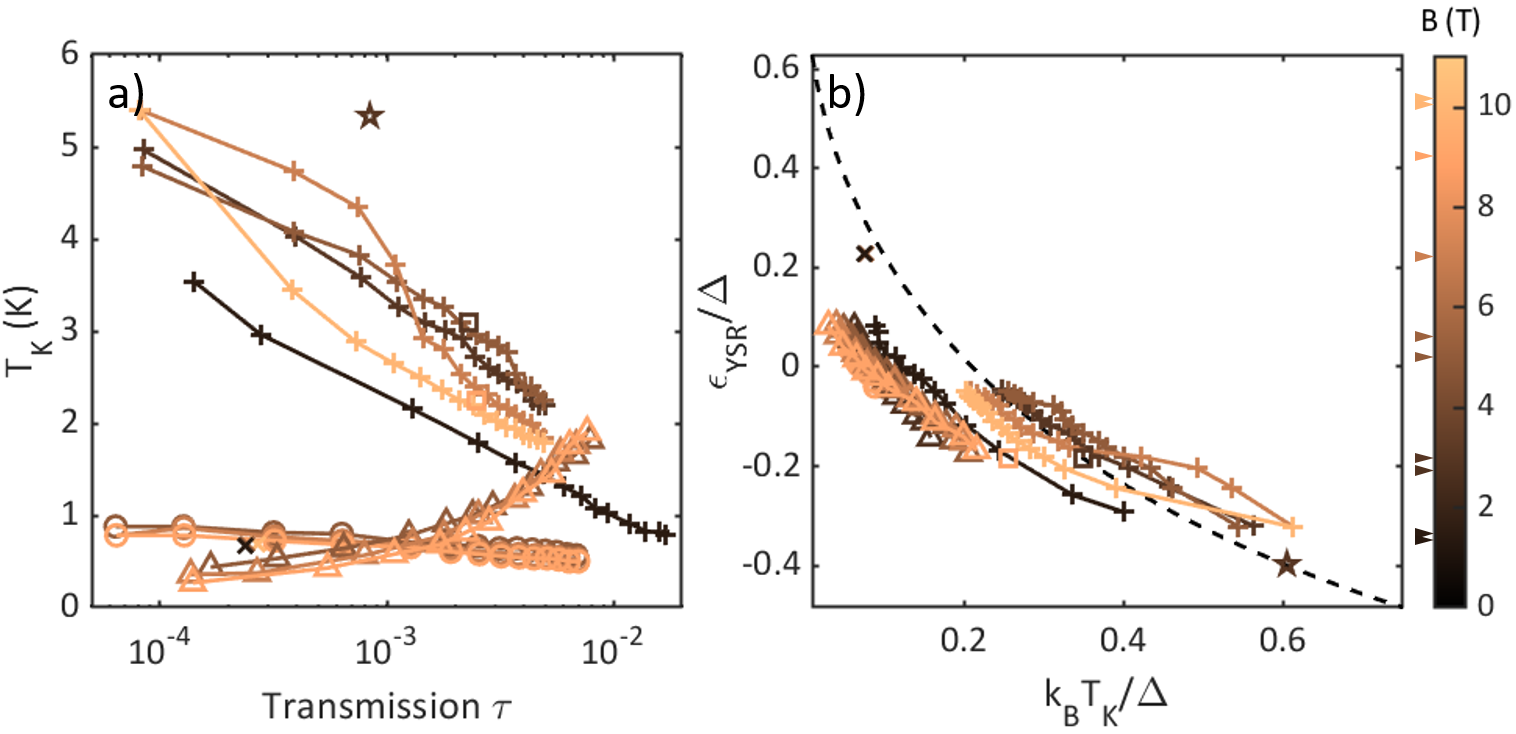}
    \caption{\textbf{The universal scaling of various YSR impurities on vanadium tip.} (a) The extracted Kondo temperature $T_\text{K}$ as a function of the transmission $\tau$. Different markers denote different impurities. The $+$ sign denotes the impurity shown in Figs.\ \ref{fig1} and \ref{fig3}. Different colors represent different magnetic fields, and the used magnetic fields are marked with arrows of corresponding colors on the left side of the colorbar. $T_\text{K}$ of the same impurity extracted from data in different magnetic fields are comparable, showing consistency throughout each dataset. (b) The universal scaling of the normalized YSR energy as a function of $k_\text{B} T_\text{K}/\Delta$. All datasets show similar tendency agreeing with the theoretical universal scaling (black dashed line).}
    \label{fig4}
\end{figure}

\textbf{Universal scaling in various YSR impurities.} Converting the fitted impurity-substrate coupling $\Gamma$ to a Kondo temperature $T_\text{K}$ through Eq.\ \eqref{eq_kondoT}, we plot $T_\text{K}$ as a function of the transmission $\tau$ in Fig.\ \ref{fig4}(a) as well as the normalized YSR energy $\varepsilon_\text{YSR}/\Delta$ as a function of $k_\text{B}T_\text{K}/\Delta$ in Fig.\ \ref{fig4}(b). The different markers (six in total: $+,\circ,\triangle,\times,\star,\Box$) label the different YSR impurities, where the YSR states represented by the three markers $+,\ \circ$, and $\triangle$ move across the QPT while the other three $\times,\star$, and $\Box$ do not move significantly when approaching the tip to the sample. The impurity shown in Figs.\ \ref{fig1} and \ref{fig3} is presented with the marker $+$, which is the same impurity as in Ref.\ \cite{karan2021superconducting}. The different magnetic fields, in which the impurities have been measured, are color coded as shown in the figure colorbar.

Depending on the interactions in the junction, the impurity-substrate coupling $\Gamma$ and concomitantly the Kondo temperature $T_\text{K}$ decreases ($+,\circ$) or increases ($\triangle$) when approaching the tip to the sample \cite{Huang2019magnetic}, indicating that the YSR state moves across the QPT from the screened spin regime to the free-spin regime or the other way around, respectively. In some cases the magnetic impurity is more rigidly bound to the tip, such that impurity is not susceptible to the atomic forces in the junction and the YSR energy does not depend on the junction transmission within the range of our measurements ($\times,\star,\Box$).

Despite the varying behavior of the impurities during tip approach, the scaling between the YSR energy and the Kondo temperature is generally universal (Fig.\ \ref{fig4}(b)). The theoretical curve predicted by the NRG theory is plotted as a dashed line. All experimental data generally reproduces the trend of the universal scaling.

\section*{Discussion}

There are multiple indications that the impurity induced YSR states in vanadium are \spinhalf\ systems ranging from single quasiparticle level tunneling and its spin selection \cite{Huang2020tunneling,huang2020spin}, to the observation of a 0-$\pi$-transition \cite{karan2021superconducting}. This is further corroborated by the observation of only one pair of YSR peaks on both sides of the QPT (see Fig.\ \ref{fig1}(b), in contrast with high spin systems \cite{Hatter2015}) and the remarkable quantitative agreement when fitting the Kondo spectra with the \spinhalf\ SIAM model using NRG theory (see Figs.\ \ref{fig3}(a) and (b)). This makes our system a model platform to study the relation between the Kondo effect and YSR states.

Similar scaling has been observed before for different kinds of impurities (not necessarily \spinhalf)  \cite{Franke2011,bauer2013microscopic,odobesko2020observation,lee2017scaling,kamlapure2019investigation,ayani2022switchable}. In most cases, the experimentally derived definition of the Kondo temperature leaves some ambiguity in the comparison with the universal scaling. Using the full microscopic NRG model we not only minimized the ambiguity but also accounted for the effect of the applied magnetic field in fitting the data. In addition, since our experimental temperature of $10\,\text{mK}$ is much smaller than the Kondo temperature, the intrinsic temperature broadening of the Kondo peak is negligible \cite{cronenwett1998tunable,otte2008role}. Furthermore, the Fermi-Dirac broadening of the probing electrode can also be neglected due to our high energy resolution at mK temperatures. Consequently, we do not need any deconvolution to extract the real Kondo signal from the experimentally broadened peaks, further achieving quantitative accuracy.

Still, the remaining deviations from the universal scaling as displayed in Fig.\ \ref{fig4}(b) raise interesting questions as to their origin. One possibility, as proposed in previous studies, is the validity of the universal scaling with respect to the parameter space of the SIAM \cite{bauer2007spectral}. For atoms on substrates in our case, both the impurity-substrate coupling $\Gamma$ and the half band width $D$ are always much lager than $\Delta$, such that the universal scaling is valid and precise independent of $U$ \cite{bauer2007spectral,kadlecova2019practical}. Therefore, this reasoning can be ruled out as outlined in the Supplementary Information \cite{supinf}. However, a realistic structured electronic band and non-constant density of states can modify the Kondo physics in a non-trivial way \cite{zitko2009num,zitko2016kondo,fernandez2019kondo} and thus go beyond the simple assumption of the universal scaling. It is therefore interesting to investigate the validity or breakdown of the universal scaling in weakly coupled systems like certain molecules and in systems with non-trivial band structure. 

We further find that the fit for different magnetic fields differs slightly even for the same impurity. This indicates that the magnetic field may change the force exerted on the impurity and in turn the impurity-substrate coupling. Also, the quenching of the superconductivity could change the atomic forces modifying the impurity-substrate coupling. This is reasonable because the relative deviation in terms of the impurity-substrate coupling is only a few percent (see Fig.\ \ref{fig3}(d)), while the resulting response in the Kondo temperature is much larger due to the exponential dependence (Eq.\ \eqref{eq_kondoT}). Other mechanisms including residual interactions with nearby spins could also influence the universal scaling behavior. 

\section*{Conclusion}

In summary, we have analyzed the scaling between the YSR energy and the Kondo temperature fits of Kondo spectra using microscopic model on the magnetic impurities at the apex of a vanadium tip. The impurities show \spinhalf\ characteristics and some of them move across the quantum phase transition during tip approach allowing for a continuous control and investigation of the scaling behavior. Directly using NRG theory in the fitting of the Kondo spectra reduces the ambiguity in extracting the Kondo temperature compared to experimentally derived definitions of the Kondo temperature. Additionally, our experimental temperature in the mK regime minimizes the experimental broadening. With this, our results corroborate the universal scaling of the NRG theory at the atomic scale with quantitative precision. The observed deviations point to a degree of freedom not captured by the current theory like modified impurity-substrate coupling in a magnetic field, which calls for further investigations. A future direction would be to apply such analysis to other \spinhalf\ systems as well as higher spin systems or coupled YSR dimers \cite{ding2021tuning}.

\section*{Acknowledgments}
We gratefully acknowledge stimulating discussions with Alfredo Levy Yeyati, Rok Žitko, Markus Ternes, and Robert Drost. This work was funded in part by the ERC Consolidator Grant AbsoluteSpin (Grant No.\ 681164) and by the Center for Integrated Quantum Science and Technology (IQ$^\textrm{\small ST}$). J.A. acknowledges funding from the BMBF within the sub-project QCOMP (Cluster4Future QSENSE). C.P. acknowledges funding from the IQ$^\textrm{\small ST}$ and the Zeiss Foundation. J.C.C. acknowledges funding from the Spanish Ministry of Science and Innovation (grant no. PID2020-114880GB-I00) and the DFG and SFB 1432 for sponsoring his stay at the University of Konstanz as a Mercator Fellow.

\clearpage
\newpage

\onecolumngrid
\begin{center}
\textbf{\large Supplementary Material for \\ Universal scaling of tunable Yu-Shiba-Rusinov states across the quantum phase transition}
\end{center}
\vspace{1cm}
\twocolumngrid

\setcounter{figure}{0}
\setcounter{table}{0}
\setcounter{equation}{0}
\renewcommand{\thefigure}{S\arabic{figure}}
\renewcommand{\thetable}{S\Roman{table}}
\renewcommand{\theequation}{S\arabic{equation}}

\vspace{0.5cm}

\section{Experiment}
The sample used in the experiments was a vanadium (100) single crystal with $>99.99$\% purity, which was prepared in ultrahigh vacuum (UHV) by multiple cycles of argon ion sputtering at around 1\,keV acceleration energy in about $10^{-6}$ mbar argon pressure followed by annealing at around 700$^\circ$C. The tip was cut from a polycrystalline vanadium wire of 99.8\% purity and subsequently prepared in UHV by Argon sputtering. To obtain a tip exhibiting clean bulk gap as well as good imaging capabilities, several rounds of field emission as well as standard tip shaping techniques were conducted on V(100) surface.

There is a sparse distribution of intrinsic magnetic impurities on V(100) surface giving rise to Yu-Shiba-Rusinov (YSR) states, which feature YSR energies spreading the whole superconducting gap, indicating flexible atomic configurations as the origin. They are \spinhalf entities likely originating from simple elements like oxygen or carbon in vanadium environments \cite{si_Huang2020tunneling,si_Huang2019magnetic,si_huang2020spin}.

When we dip the tip onto the surface, there is a finite possibility that YSR states are introduced onto the tip apex, which we call a YSR tip \cite{si_Huang2020tunneling,si_huang2020spin,si_karan2021superconducting}. A computer program was written to automatically control such a tip shaping process and to obtain a YSR state with desired properties (for example, YSR energy, peak intensity and peak asymmetry). The YSR states on the tip share similar statistics as the intrinsic sample YSR states being also flexible in energy and asymmetry \cite{si_Huang2020tunneling}, but they are much more controllable since we can fabricate them with desired properties automatically. 

One intriguing aspect regarding YSR tips is that they feature a YSR state that moves across the quantum phase transition (QPT) during tip approach \cite{si_karan2021superconducting}. Notice that the same kind of YSR impurities also exist intrinsically on the surface, but due to their sparseness, it is much more practical to fabricate them on purpose on the tip apex. In general, the YSR states will move in energy during tip approach, because the changing atomic force will more or less pull (or push) the impurity away (or towards) the superconducting host, resulting in a changing coupling and concomitantly changing YSR energy. Nevertheless, the extent of this effect depends on the specific situation of each impurity. Some impurities are more softly bound and thus more prone to the influence of the atomic forces in the junction, and the corresponding YSR energy moves significantly as a function of transmission (or conductance). If the exchange interaction happens to be in the correct range, the YSR energy can move across the QPT. Other impurities are more rigidly bound and therefore less affected by the atomic forces across the junction, where the corresponding YSR energy moves only slightly or stays constant with changing transmission. In the main text of this paper, we have presented both scenarios.

\section{Numerical renormalization group theory}
We used the “NRG Ljubljana” package for NRG simulations under the framework of the single impurity Anderson model (SIAM) \cite{si_pruschke2009energy}. In the code, all energy scales are in the unit of the half bandwidth $D$ of the bulk conduction electrons, which is arbitrary. To be able to compare with the experiment directly, we fixed the superconducting order parameter $\Delta=750\times10^{-6}$ in the simulations such that all energy parameters are in the unit of eV, which becomes experimentally relevant (equivalent to setting $D$ to 1\,eV). All energy scales in this supplementary information are also in unit of $D$, if not specifically discussed otherwise.

In all NRG claculations, we used the renormalization parameter $\Lambda= 2$, the discretization scheme from Ref. \cite{si_pruschke2009energy}, and the z-averaging technique with $N_z = 12$ \cite{si_zitko2009numerical}. The Fermi level was chosen to be at zero energy.

\section{The single impurity Anderson model in the small \textit{U} and the large \textit{U} limit}

As discussed in the main text, the Kondo temperature can be expressed in the terms of the SIAM parameters \cite{si_schrieffer1966relation,si_kadlecova2019practical} as 
\begin{equation}
    k_\text{B}T_\text{K} = D_\text{eff}\sqrt{\rho J}\exp\left(-\frac{1}{\rho J}\right)\text{, with } \rho J=\frac{8\Gamma}{\pi U}\frac{1}{1-4(\delta/U)^2},
\label{eq1}
\end{equation}
where the effective band width $D_\text{eff}=0.182U\sqrt{1-4(\delta/U)^2}$ for $U\ll 1$ and $D_\text{eff}$ is a constant on the order of 1 for $U\gg 1$ \cite{si_Yoshioka2000,si_zitko2007many} (in the NRG implementation used in this paper $D_\text{eff}$ is 0.5 for $U\gg1$).

\begin{figure}
    \centering
    \includegraphics[width=\columnwidth]{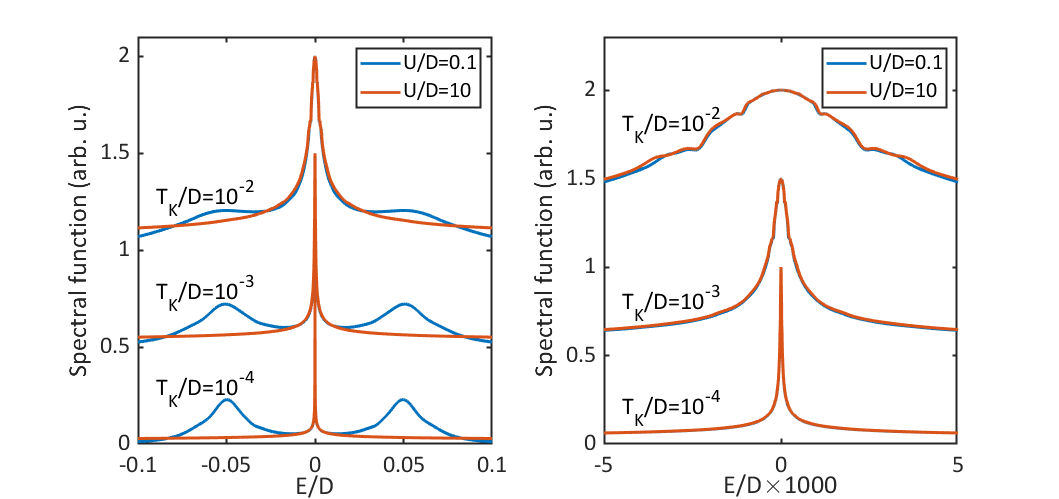}
    \caption{\textbf{Spectral functions of SIAM simulated with NRG in the symmetric case ($\delta=0$).} Curves for different Kondo temperature are offset for clarity (blue: $U\ll1$ and red: $U\gg1$). The corresponding $\Gamma$ of a given Kondo temperature used in the simulation is obtained from Eq.\ \eqref{eq1} for a given $U$. For $U=0.1$, the impurity levels near $\pm U/2=\pm 0.05$ appear inside the plotted energy range. The right panel is a zoom-in to the left panel showing the Kondo peaks around zero energy confirming $U$ to be irrelevant.}
    \label{fig3}
\end{figure}

\begin{figure}
    \centering
    \includegraphics[width=\columnwidth]{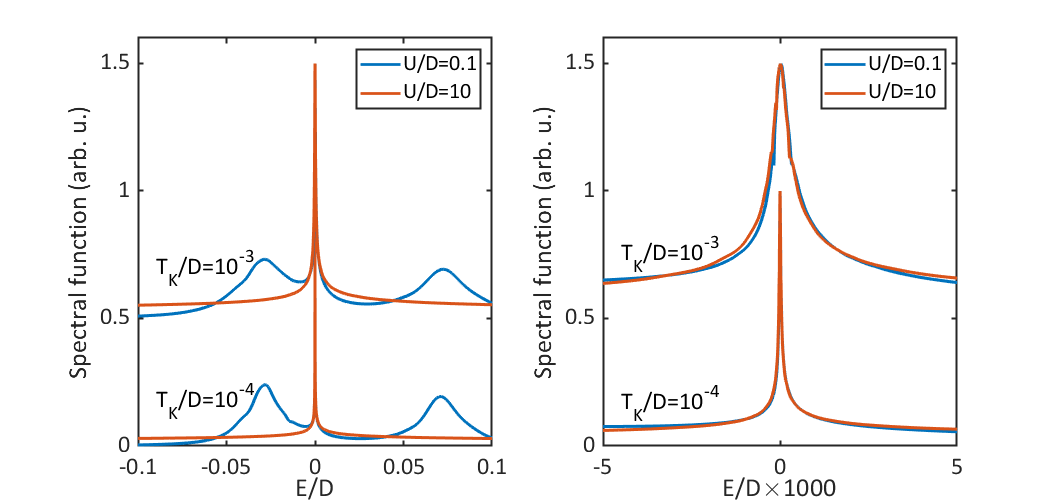}
    \caption{\textbf{Spectral functions of SIAM simulated with NRG in the asymmetric case ($\delta=0.2U$).} Curves for different Kondo temperature are offset for clarity (blue: $U\ll1$ and red: $U\gg1$). The corresponding $\Gamma$ of a given Kondo temperature used in the simulation is obtained from Eq.\ \eqref{eq1} for a given $U$. For $U=0.1$, the impurity levels near $\delta\pm U/2=-0.03,0.07$ appear inside the plotted energy range. The right panel is a zoom-in to the left panel showing the Kondo peaks around zero energy confirming $U$ to be irrelevant.}
    \label{fig4}
\end{figure}

As discussed in the main text, not all SIAM parameters ($U, \delta, \Gamma$) are equally relevant for the low energy physics (Kondo and YSR, for example), which is dominated by the Kondo temperature. $\delta$ is responsible for the asymmetry of the spectra, but $U$ is free to choose as it has no significant effect on the Kondo or Shiba spectra. This is shown in Figs. \ref{fig3} and \ref{fig4}. 

In Fig.\ \ref{fig3}, the simulated spectra for $U=0.1$ and $U=10$ with $\Gamma$ corresponding to certain Kondo temperatures (using Eq.\ \eqref{eq1}) are plotted as blue and red curves respectively, with different Kondo temperature ($T_\text{K}=10^{-2},10^{-3},10^{-4}$) offset for clarity. With increasing Kondo temperature, the Kondo peaks near Fermi energy broaden, and the impurity levels near $\pm\frac{U}{2}$ (only visible here for $U=0.1$) also broaden due to increasing $\Gamma$. The right panel is a zoom-in to the Kondo peak near the Fermi energy in the left panel, showing no difference for different $U$.

The asymmetric situation is shown in Fig.\ \ref{fig4} ($\delta=0.2U$). With increasing Kondo temperature, the Kondo peaks as well as the impurity levels broaden. One difference compared to $\delta=0$ is that the impurity levels no longer exist symmetrically around zero energy but appear at around $\delta\pm\frac{U}{2}$ (only visible here for $U=0.1$) and the resulting Kondo peak is also asymmetric. In the zoom-in to the Kondo peak in the right panel of Fig.\ \ref{fig4}, the independence of the spectra over $U$ still holds. The slight deviation comes from the close vicinity of the impurity peaks to zero energy for $U=0.1$, whose shoulder adds some spectral weight to the Kondo peak and reshapes it. Therefore, the choice of $U$ is irrelevant for fitting experimental Kondo spectra, but to avoid confusion of nearby impurity levels from using small $U$ (experimentally we also never observe such levels in the spectra), we fix $U=10$ in the main text of this paper.

\section{The validity of the universal scaling}

We consider the condition for SIAM parameters at the QPT according to the universal scaling $k_\text{B}T_\text{K}/\Delta=\alpha$, where $\alpha=0.24$. Other works find values between 0.2 and 0.3 depending on the NRG implementation \cite{si_Yoshioka2000,si_bauer2007spectral,si_bulla2008numerical,si_kadlecova2019practical}. For simplicity, we only consider the symmetric case in the following discussion ($\delta=0$). For $U\ll 1$, the QPT happens when 
\begin{equation}
    k_\text{B}T_\text{K} = 0.182U\sqrt{\frac{8\Gamma}{\pi U}}\exp\left(-\frac{\pi U}{8\Gamma}\right)=\alpha\Delta, 
\end{equation}
and thus 
\begin{equation}
    \frac{\Delta}{U} = \frac{0.182}{\alpha}\sqrt{\frac{8\Gamma}{\pi U}}\exp\left(-\frac{\pi U}{8\Gamma}\right) = \frac{\Delta}{\Gamma}\frac{\Gamma}{U}
    \label{eq:qptline}
\end{equation}
is a function of $\frac{U}{\Gamma}$. Additionally, since $\frac{\Delta}{\Gamma}=\frac{\Delta}{U}\frac{U}{\Gamma}$, $\frac{\Delta}{\Gamma}$ is also a function of $\frac{U}{\Gamma}$. Consequently, if we use either $\frac{\Delta}{U}$ or $\frac{\Delta}{\Gamma}$ as the $y$-axis and $\frac{U}{\Gamma}$ as the axis (or their inverse), we can plot the QPT line in Eq.\ \eqref{eq:qptline} as expected from the universal scaling (solid line in the left panel of Fig.\ \ref{fig2}).

For $U\gg 1$, the plot would be different. There, the QPT happens when 
\begin{equation}
    k_\text{B}T_\text{K} = D_\text{eff}\sqrt{\frac{8\Gamma}{\pi U}}\exp\left(-\frac{\pi U}{8\Gamma}\right)=\alpha\Delta,
\end{equation}
so $\Delta$ is directly a function of $\frac{U}{\Gamma}$ and the $y$-axis of the plot should be changed to $\Delta$ (actually it is $\frac{\Delta}{D}$ because we reference all energies to the half bandwidth $D$). This is plotted as solid line in the right panel of Fig.\ \ref{fig2}.

Furthermore, we can simulate the YSR energy and find out the point of the QPT using NRG, which is plotted as circles in Figs. \ref{fig2}. For $\Gamma\ll U$, the NRG prediction obeys the universal scaling quite well, which is consistent with discussions in the literature \cite{si_bauer2007spectral,si_kadlecova2019practical}. In STM, we always have $\Delta\ll\Gamma$ and $\Delta\ll D$, therefore the universal scaling holds independent of $U$.

\begin{figure}
    \centering
    \includegraphics[width=\columnwidth]{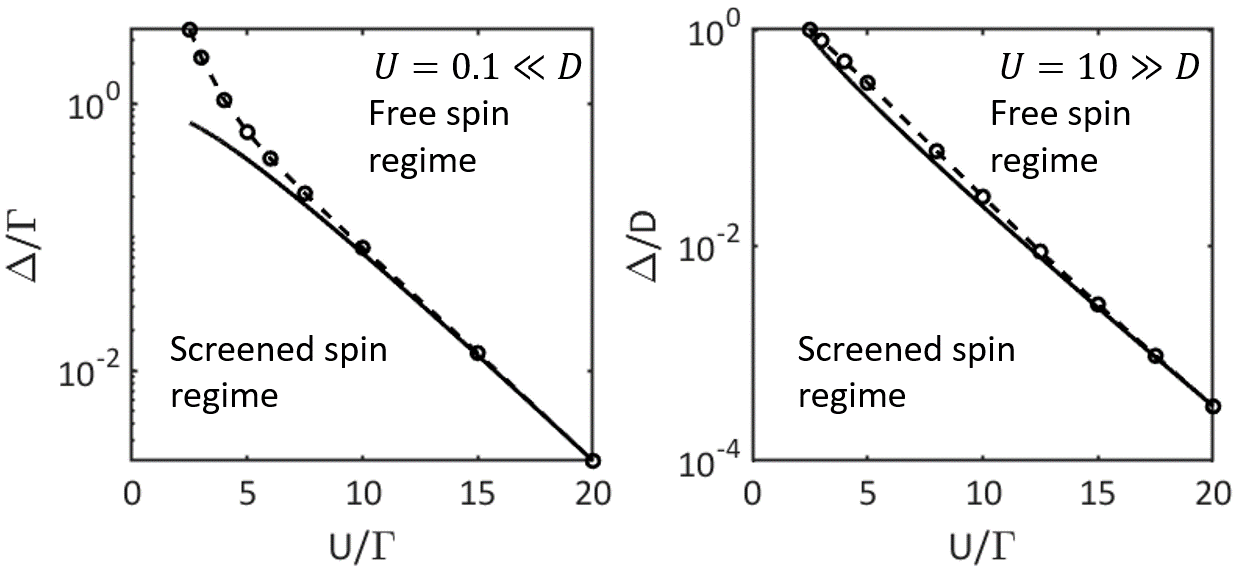}
    \caption{\textbf{Validity of the universal scaling in the SIAM in small $U$ and large $U$ limit.} The circles represent data points obtained using NRG simulations with $U=0.1\ll1$ (left panel) and $U=10\gg1$ (right panel) with various $\Delta$ and $\Gamma$. The solid black lines are expectation from the universal scaling $k_\text{B}T_\text{K}=\alpha\Delta$ (for details see text).}
    \label{fig2}
\end{figure}

\section{Definitions of Kondo temperatures}
One definition of the Kondo temperature widely adopted in experiments is the half-width-half-maximum (HWHM) of the Kondo peak $T_\text{K,exp}=\Gamma_\text{HWHM}/k_\text{B}$ at absolute zero temperature \cite{si_Ternes2009,si_gruber2018kondo}, which can be extended to Kondo peaks at finite temperatures via the Fermi liquid theory \cite{si_otte2008role,si_Ternes2009,si_gruber2018kondo}.

Usually the observed Kondo spectra feature more complex shapes and require a fitting procedure to extract the Kondo temperature, and the most common phenomenological fitting functions are the Fano function \cite{si_fano1961effects,si_nagaoka2002temperature,si_otte2008role,si_Ternes2009} 
\begin{equation}
    \rho_\text{Fano}(E)\propto\frac{(q+\epsilon)^2}{1+\epsilon^2},
\end{equation}
where $\epsilon=\frac{E-E_0}{\Gamma_\text{Fano}}$ and the Frota function \cite{si_frota1992shape,si_zitko2009splitting,si_pruser2012mapping,si_von2015spin,si_gruber2018kondo} 
\begin{align}
    \rho_\text{Frota}(E)&\propto \text{Im}\left(ie^{i\phi}\sqrt{\frac{i\Gamma_\text{Frota}}{E-E_0+i\Gamma_\text{Frota}}}\right)=\\
    &=\left(1+\epsilon^2\right)^{-1/4}\cos{\left(\phi+\frac{1}{2}\arctan{\epsilon}\right)},
\end{align}
where $\epsilon=\frac{E-E_0}{\Gamma_\text{Frota}}$.

The parameters extracted from these two fitting methods are mathematically related to the HWHM by $\Gamma_\text{HWHM}=\Gamma_\text{Fano}=2.54246\Gamma_\text{Frota}$ \cite{si_gruber2018kondo,si_ternes2017probing}. Notice, however, the HWHM of the Fano and Frota functions obey the above relation only for the symmetric case ($q\rightarrow\infty$ for the Fano case and $\phi=0$ for the Frota case).

Apart from possible discrepancies arising from different methods of extracting the Kondo temperature from measured spectra, even theoretically various definitions of Kondo temperature exist because the Kondo effect is not a sharp phase transition like superconductivity but rather a smooth transition \cite{si_wilson1975renormalization,si_nozieres1974fermi,si_bulla2008numerical,si_hewson2004renormalized,si_hamann1967new,si_nagaoka1965self}. As a result, a deviation up to a factor of four can occur \cite{si_bauer2013microscopic} (see Fig.\ \ref{fig1} for a comparison between HWHM and the Kondo temperature from Wilson's definition \cite{si_wilson1975renormalization}). In this paper, we consistently use Wilson's definition of the Kondo temperature which is commonly used in NRG literatures \cite{si_bulla2008numerical,si_kadlecova2019practical}, so that such ambiguity is avoided.

\begin{figure}
    \centering
    \includegraphics[width=\columnwidth]{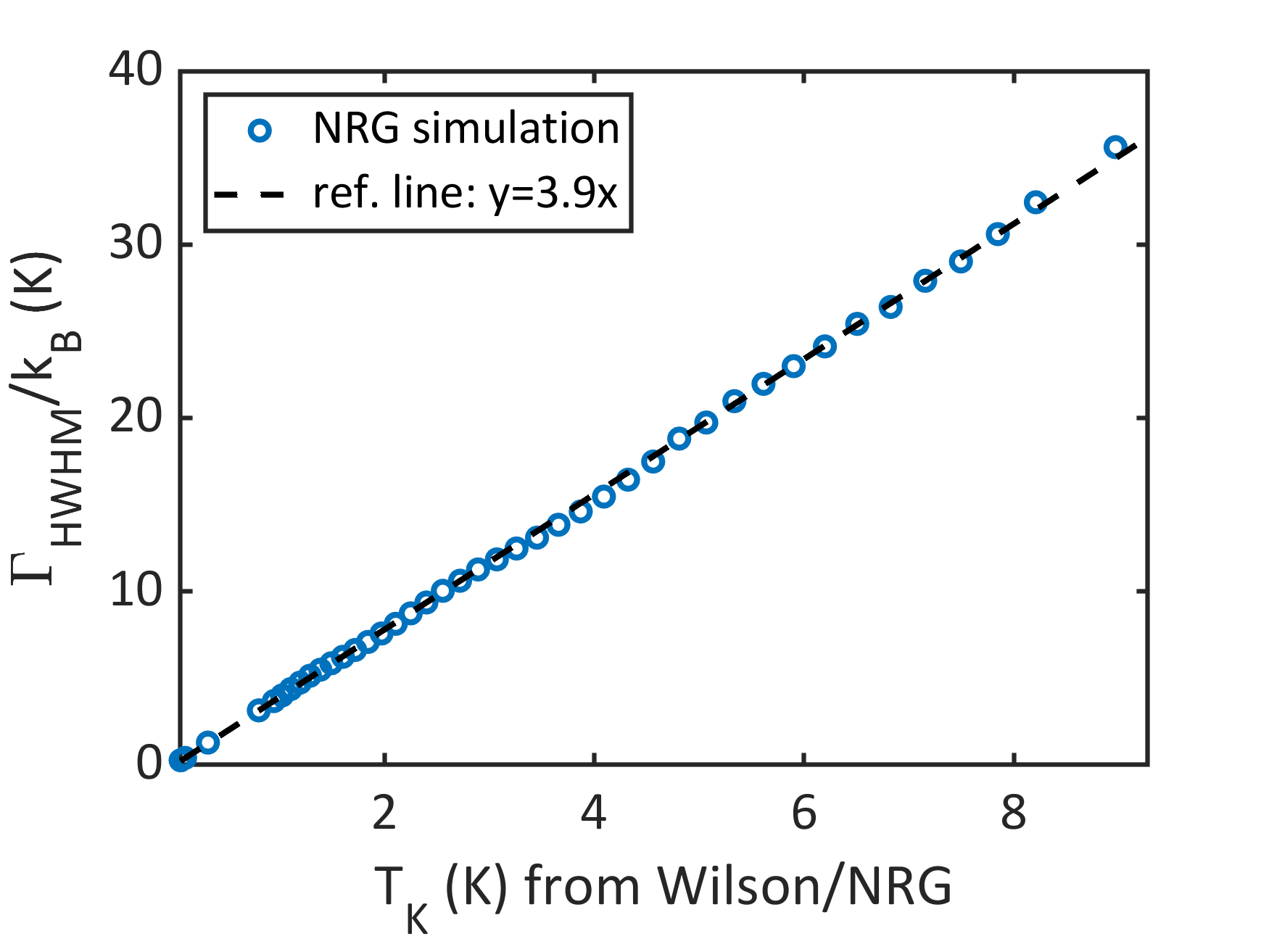}
    \caption{\textbf{Comparison of different definitions of Kondo temperature.} The spectral functions are simulated using the NRG theory within symmetric SIAM for various $\Gamma$ (with $U=10, \delta=0$). For each spectrum, the Kondo temperature $T_\text{K}$ is calculated using the SIAM parameters from the Wilson's definition commonly used in the literature (also used in the main text) \cite{si_wilson1975renormalization}. The HWHM $\Gamma_\text{HWHM}$ is determined from the spectra directly and plotted against $T_\text{K}$. A straight line through the origin that best fit the points is displayed, giving an estimation of $\Gamma_\text{HWHM}\approx3.9k_\text{B}T_\text{K}$, showing a factor of nearly four between the common experimental definition of $T_\text{K,exp}$ based on the HWHM and $T_\text{K}$ from the NRG formalism. The conversion constant is consistent with the literature \cite{si_pruschke2009energy,si_pruser2012mapping}.}
    \label{fig1}
\end{figure}

\section{Adding Fano process in NRG formalism}
The SIAM contains intrinsic electron-hole asymmetry resulting in the asymmetry of the spectral function. Another widely accepted mechanism of the asymmetry is the Fano process. In the main text we attributed the asymmetry to the intrinsic one from the SIAM, here we discuss the alternative.

\begin{figure}
    \centering
    \includegraphics[width=\columnwidth]{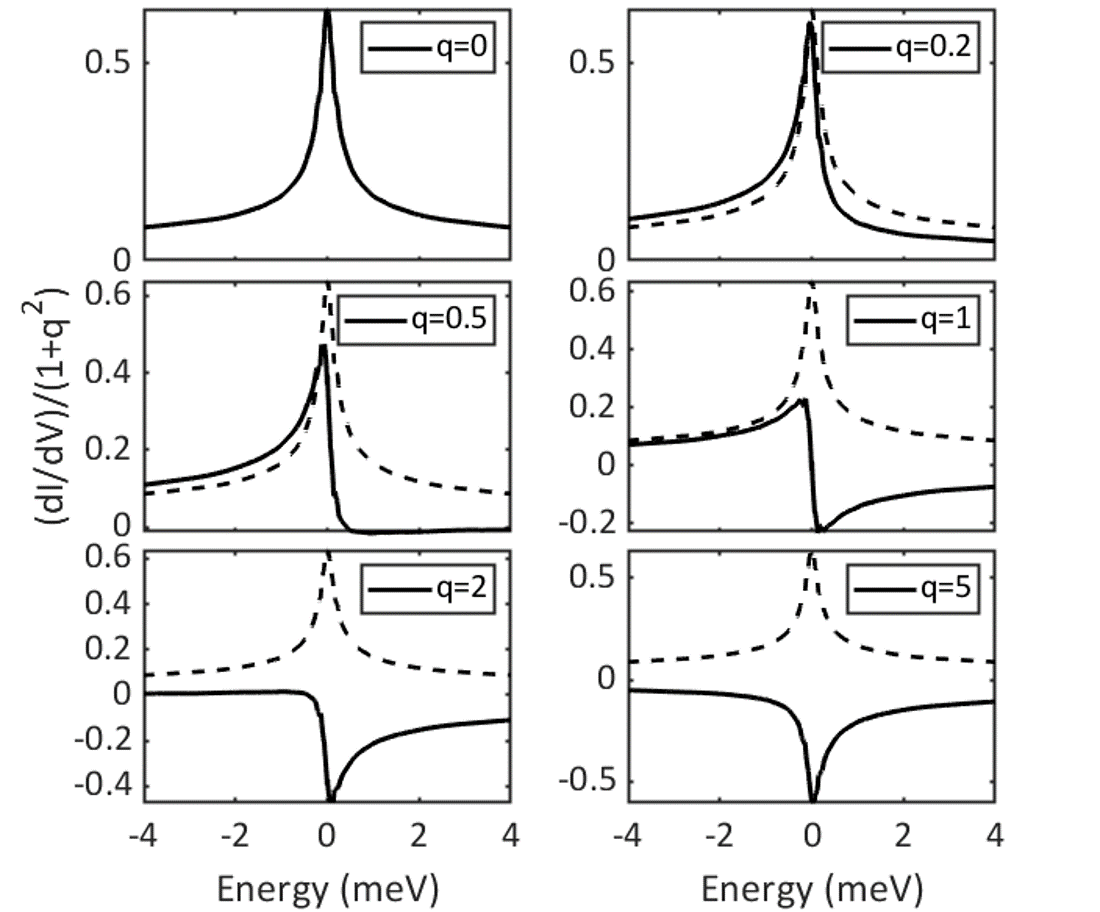}
    \caption{\textbf{Simulated dI/dV from NRG considering Fano process with different $q$.} The dashed lines are the same NRG simulated Kondo spectra with parameters $\delta=0, \Gamma=0.5, B=0$, all units in $D=1\,\text{eV}$. The solid lines are after consideration of the Fano process using Eq.\ \eqref{eq_fano_3} with different $q$ displayed in the corresponding legend. For $q=0$, Fano process is absent and two lines overlap. With increasing $|q|$, the $dI/dV$ becomes more and more asymmetric and then becomes a dip in the end. Changing the sign of $q$ will reverse the asymmetry into the other direction.}
    \label{fig5}
\end{figure}

\begin{figure}
    \centering
    \includegraphics[width=\columnwidth]{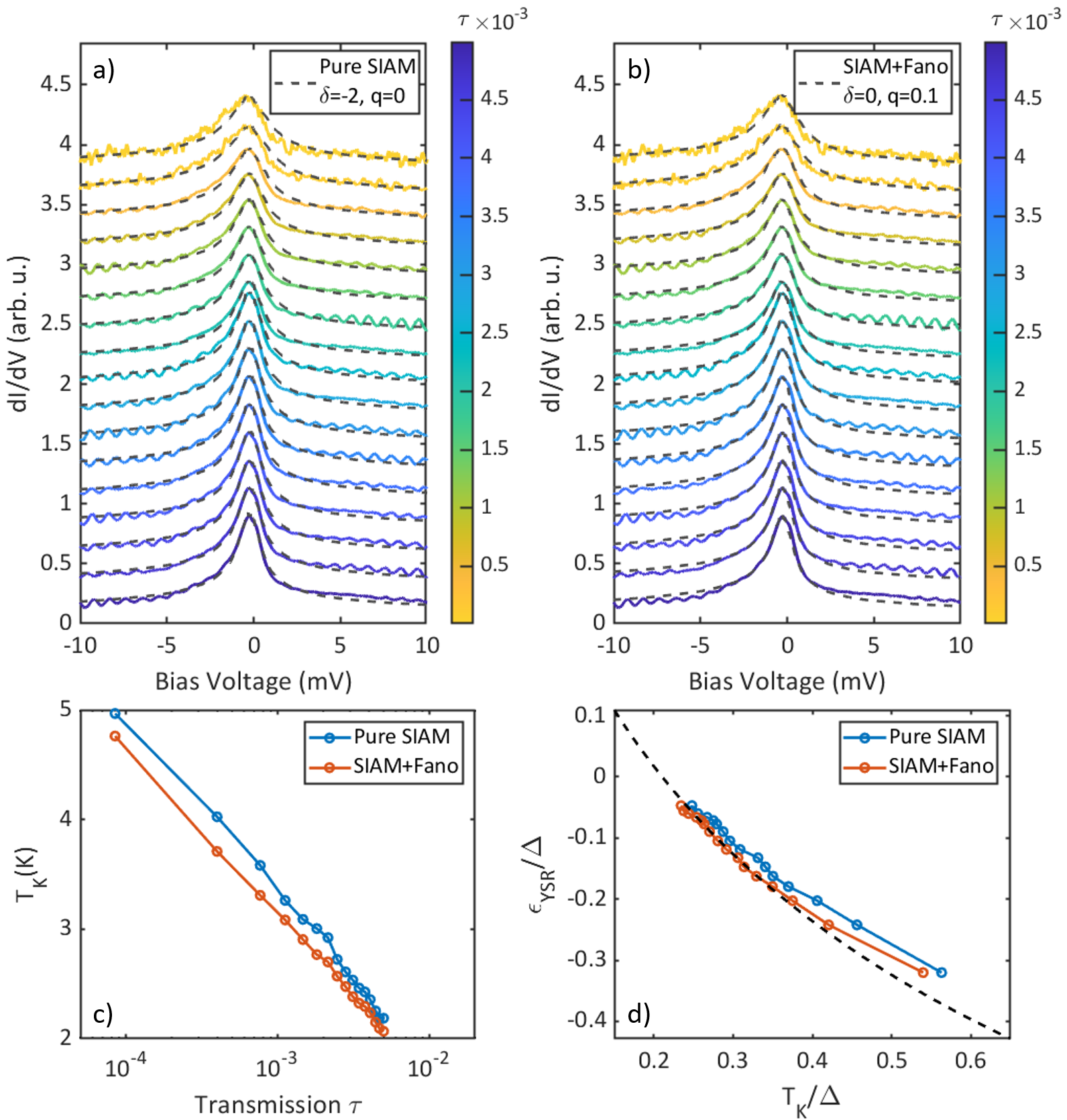}
    \caption{\textbf{Comparison of fitting using pure SIAM and fitting considering Fano processes.} Data is the $B=3\,\text{T}$ one from the main text. a) Fitting considering asymmetry from pure SIAM with no Fano ($\delta=-2,q=0$). b) Fitting considering asymmetry from Fano keeping SIAM symmetric ($\delta=0,q=0.1$). c) Extracted Kondo temperature $T_\text{K}$ as a function of normal state trasmission $\tau$. d) Normalized YSR energy as a function of $T_\text{K}/\Delta$ showing little difference from the two fitting procedures, indicating robustness of the result in the main text.}
    \label{fig6}
\end{figure}

To incorporate the Fano process into the simulated spectra from NRG formalism, we need to use the Green's function theory. The spectral function
\begin{equation}
   A(\omega)=-\frac{1}{\pi}\Im G^r(\omega)
\label{eq_fano_1}
\end{equation}
is directly obtained from NRG simulation. The real part of the Green's function can be obtained through the Kramers Kronig relations
\begin{equation}
    \Re G^r(\omega) = -\frac{1}{\pi}\text{P}\int_{-\infty}^{\infty}d\omega'\frac{\Im G^r(\omega')}{\omega-\omega'},
\label{eq_fano_2}
\end{equation}
where $\text{P}$ stands for the principal value integral. Then the final $dI/dV$ modified by the Fano process can be written as \cite{si_ujsaghy2000,si_zitko2011kondo}
\begin{equation}
    \frac{dI}{dV}\big|_\text{Fano} \propto -\frac{1}{\pi}\left[(1-q^2)\Im G^r(\omega)+2q\Re G^r(\omega)\right],
\label{eq_fano_3}
\end{equation}
with $q$ being the parameter controlling the extent of Fano lineshape. When $q=0$, Eq.\ \eqref{eq_fano_3} reduces to Eq.\ \eqref{eq_fano_1}, recovering to the original spectral function. The effect of $q$ on the NRG simulated Kondo spectra is shown in Fig.\ \ref{fig5}. It can be seen in conjunction with the experimental data in Fig.\ \ref{fig6} that we only have a very small $q$, indicating that the Fano process is weak.

We can now fit again $B=3\,\text{T}$ data shown in the main text assuming all asymmetry stemming from Fano lineshape in Eq.\ \eqref{eq_fano_3} (and assuming $\delta=0$ in the SIAM), shown Fig.\ \ref{fig6}(b). The fit using asymmetric SIAM without Fano in the main text is reproduced here in Fig.\ \ref{fig6}(a) for better comparison. Both fits are quite good and can hardly be distinguished which one is better. The Kondo temperature $T_\text{K}$ extracted are also very similar (Figs.\ \ref{fig6}(c) and (d)), with a deviation below around 4\%. In the example here the scaling is close to the universal one (Fig.\ \ref{fig6}(d)), but generally it can be seen that the extracted $T_\text{K}$ is robust irrespective of the fitting in our case. The reason might be that our Kondo spectra feature prominent peaks with only a bit asymmetry, and the Fano process, if exists at all, is very weak.

\end{document}